\documentclass[fleqn,usenatbib,useAMS]{mnras}
\usepackage{xfrac}
\usepackage{xcolor}
\usepackage{array}
\usepackage{amsmath,amssymb,amsfonts}
\usepackage{latexsym}
\usepackage{epstopdf} 
\usepackage{wasysym}
\usepackage{float}
\usepackage{array,multirow}
\usepackage{booktabs}
\usepackage{geometry}
\usepackage[strict]{changepage}
\usepackage{subfloat}
\usepackage{mathptmx}

\usepackage[T1]{fontenc}

\newcommand{\appropto}{\mathrel{\vcenter{
			\offinterlineskip\halign{\hfil$##$\cr
				\propto\cr\noalign{\kern2pt}\sim\cr\noalign{\kern-2pt}}}}}

\title[Enrichment in Nova]{The Effect of Enriched Accreted Matter on the Development of Novae}

\author[Y. Hillman and M. Gerbi]{
	Yael Hillman$^{1}$\thanks{Yael Hillman e-mail: yaelhi@ariel.ac.il} and Maya Gerbi$^{1}$	\\
	$^{1}$Department of Physics, Ariel University, Ariel, POB 3, 4070000, Israel\\
}

\date{Accepted XXX. Received YYY; in original form ZZZ}

\pubyear{2022}
\begin{document}
	\label{firstpage}
	\pagerange{\pageref{firstpage}--\pageref{lastpage}}
	\maketitle
	
	\begin{abstract}
	The development of a nova eruption is well known to be determined by the white dwarf (WD) mass and the rate at which it accretes mass from its donor. One of the advancements in this field is the understanding that the occurrence of a nova eruption depends on the presence of heavy elements in the envelope, and that the concentration of these elements is highly dependent on the time allotted for accretion. This results in many features of the eruption being correlated with the mass fractions of heavy elements in the ejected material, however, the accreted material is always assumed to be of solar metallicity. Here we explored the entire range of accretion rates onto a 1.25$M_\odot$ WD for two cases of highly enriched accreted material and find enrichment to have an influence on certain features for high accretion rates, while the effect of enrichment on low accretion rates is negligible. We further find that the ignition of the thermonuclear runaway which is known to be dependent on the accumulation of a critical mass, is actually dependent on the accumulation of a critical amount of heavy elements. 
	\end{abstract}
	

	\section{Introduction}\label{sec:intro}
Novae are thermonuclear runaways (TNR) that occur on the surfaces of white dwarfs (WD). They are the inevitable result of non-degenerate matter being pulled from the envelope of a close companion which is over filling its Roche-lobe (RLOF), and accumulating on the degenerate surface of the WD \cite[]{Starrfield1972}. The temperature at the base of the accreted layer increases due to pressure build up, eventually becoming hot enough for burning hydrogen into helium via the CNO cycle \cite[]{Prialnik1984}.  

In the CNO cycle, carbon (C), nitrogen (N) and oxygen (O) serve as catalysts in the fusion of four hydrogen atoms into one atom of helium \cite[e.g.,][]{Kenyon1983,Gehrz1998,Jose1998}. This implies that higher abundances of these elements, should accelerate the fusion process, changing the nature of the nova cycle. However, since the process leading to a nova eruption is complex, involving several stages, including the production and beta-decay of a handful of isotopes, as well as being dependent on multiple parameters, it cannot be assumed that the fusion rate will simply be correlated linearly with the abundances of C, N and O. Thus, the resulting outcome of a nova eruption due to enriched accreted material cannot be directly predicted solely based on the abundances \cite[]{Wiescher1999}.

Few authors have directly tested the  influence of different abundances on the outcoming nova. For example, \cite{Starrfield1974a} carried out multiple calculations of novae, varying the abundance of C, N and O in the envelope, while assuming arbitrary compositions and atoning the differences to be possible outcomes of contamination from the core. \cite{Kenyon1983} simulated nova eruption on a 0.8$M_\odot$ WD with envelopes of solar composition, and enriched with 20$\%$ heavy elements (roughly seven times solar enrichment), and obtained differences between the two calculations, of a few to a few tens of percent, at the onset of the TNR, in various parameters, such as the luminosity, effective temperature and the critical mass.

The envelope of the WD may become enriched from within during the accretion phase, as a result of diffusive and/or convective mixing and/or dredge up from the upper layers of the degenerate core adjacent to the accreted envelope \cite[e.g.,][]{Prikov1995,Glasner1997}. The extent of envelope enrichment due to these processes is a consequence of the accretion rate ($\dot{M}$) because $\dot{M}$ is a key parameter in setting the total accretion time. Therefore, since a lower accretion rate dictates a longer accretion phase, lower rates will allow more time for mixing material from the core into the accreting envelope and visa versa. This means that for lower accretion rates the WD envelope will have a higher abundance of heavy elements, thus more catalysts to fuse the hydrogen. The many authors that have explored (via modeling) the affect of the different input parameters on the outcome of a nova \cite[e.g.,][]{Starrfield1971,Prialnik1992, Prikov1995,Yaron2005,Epelstain2007,Denissenkov2013,Hillman2020,Hillman2021a} and have determined that the WD mass and accretion rate are key parameters in determining the outcome of nova eruptions, have all assumed a solar type donor, thus, their resulting envelope enrichment is due solely to mixing from the core, and not external.

The assumption of solar composition accreted material is an excellent one when the purpose is to simulate nova eruptions that occur in cataclysmic variables (CVs), i.e., systems with a low mass main sequence donor - a red dwarf (RD). However assuming solar composition accreted material for simulations of systems hosting an evolved donor, such as a red giant (RG) or a star on the asymptotic giant branch (AGB), which may have enriched envelopes might be a less accurate assumption. This is because when a star reaches the RG branch (RGB) it has exhausted the hydrogen in its core, it expands, and becomes more convective. On the AGB the star additionally experiences thermal pulses which enhances the mixing even more. These mixing processes can alter the envelope composition which may result in higher heavy element abundance  \cite[e.g.,][]{Prantzos1996,vanLoon2005,Ziurys2006} for different initial stellar masses and/or at different epochs.

To date, the combined influence of the accretion rate and the enrichment of the accreted material on the outcome of a nova has not been explored. 
To assess the possibility of a classical (or recurrent) nova occurring in a system with a giant donor, \cite{Hillman2021} have compared between the behavior of nova eruptions on the surface of a $1.25M_\odot$ WD in a CV system and in a wide binary, where the accretion is from the wind of an AGB donor (both models accreting solar composition material). They found that for epochs of similar average accretion rates, the behavior of the two models is indistinguishable. This demonstrates that for a given WD mass, it is not a priori the type of system that determines the outcome, but the accretion rate. 
They have included a preliminary test using more enriched accreted material, and found minor differences, stressing that in order to determine if their conclusions are robust, the test must be expanded to additional models.
 
In this study, we have carried out simulations of novae on a carbon-oxygen WD of $1.25M_\odot$ accreting matter with different enrichments for the entire range of plausible constant accretion rates. In the following section we describe the model, the range of parameters and our computational method, in \S \ref{sec:results}, we compare the results of the different simulations followed by a discussion and our conclusions in \S \ref{sec:conclusions}.

\section{Models and computational method}\label{sec:models}
The simulations were performed using a hydrodynamic Lagrangian stellar evolution code described thoroughly in \cite{Prikov1995,Yaron2005,Epelstain2007} and \cite{Hillman2015}. We performed a series of 18 simulations of multiple consecutive novae eruptions, using the same carbon-oxygen (equal parts) WD model with a mass of 1.25$M_\odot$ for all of the simulations, changing only two input parameters: (1) the accretion rate ($\dot{M}$), which was taken as $10^{-7}$, $10^{-8}$, $10^{-9}$, $10^{-10}$, $10^{-11}$ and $10^{-12}M_\odot yr^{-1}$; and (2) the metallicity of the accreted matter ($Z$), which was taken as solar composition ($Z=Z_\odot$), enriched to twice the solar abundance ($Z=2Z_\odot$) and enriched to five times the solar abundance ($Z=5Z_\odot$).
The combinations of these parameters yields 18 variations that were allowed to run for a few tens of consecutive nova cycles of accretion and eruption, while recording for each cycle the accreted and ejected masses ($m_{\rm{acc}}$ and $m_{\rm{ej}}$ respectively), the composition of the envelope  and of the ejected mass, the maximum temperature attained during an eruption ($T\rm_{max}$) as well as the surface temperature ($T\rm_{eff}$) and the temperature in the core of the WD ($T_c$), the time between two consecutive eruptions ($t\rm_{rec}$), the amplitude of the eruption ($A\rm_{bol}$) and the kinetic and radiative energy ($E\rm_{kin}$ and $E\rm_{rad}$ respectively). All the data presented here are averaged over multiple cycles in order to ensure typical values.

\section{Results}\label{sec:results} 
We find a general correlation between the enrichment and the critical mass needed for triggering a TNR - a more enriched envelope requires less mass. The need for less mass entails differences in most of the other features as well, while the accretion rate, in addition to having a larger influence on its own, also plays a key roll in the \textit{extent} of the difference \textit{caused} by enrichment, as will now be explained in detail.

\begin{figure}
	\begin{center}
		{\includegraphics[trim={0.55cm 1.9cm 0.45cm 0.5cm},clip ,
			width=0.99\columnwidth]{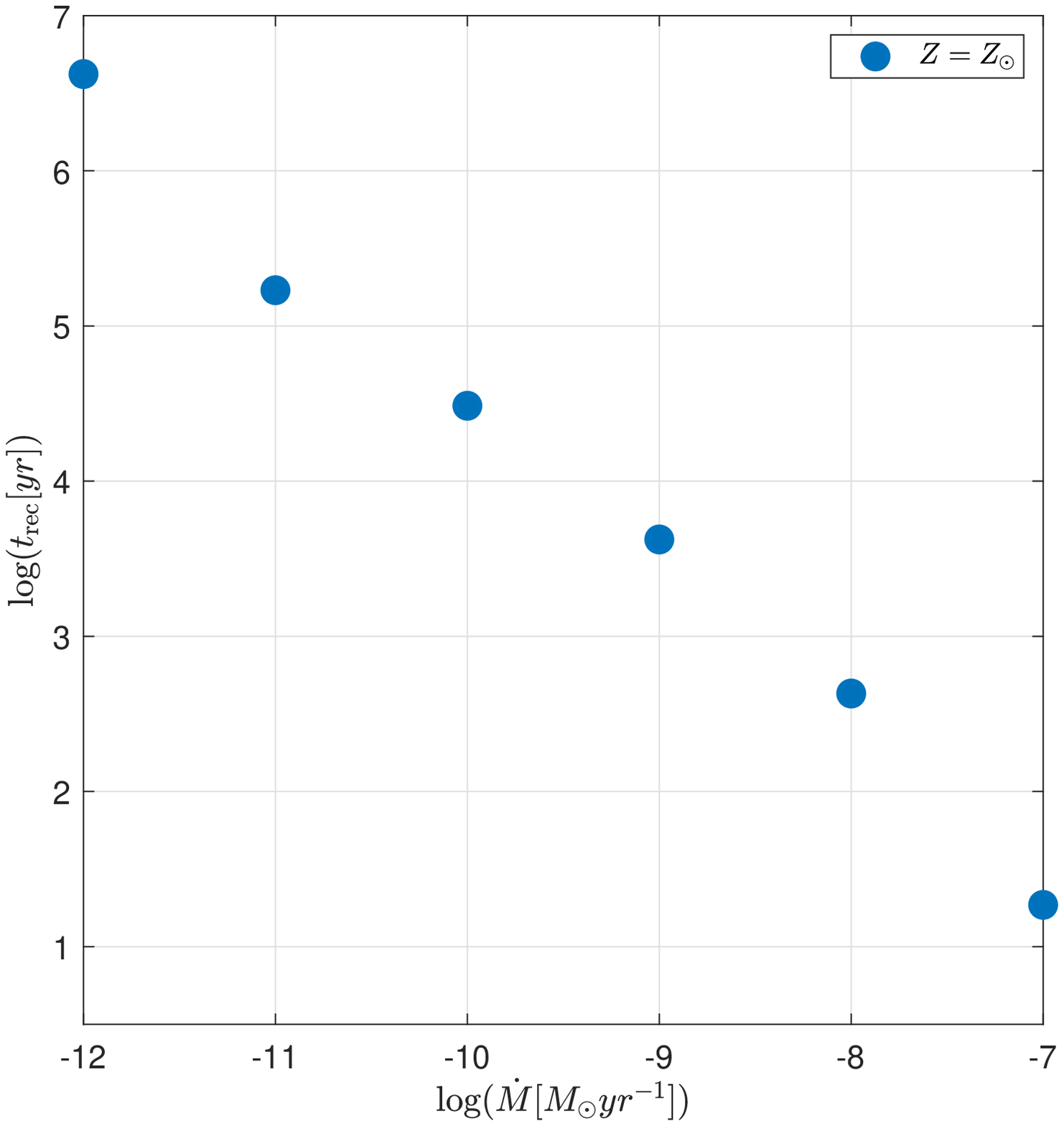}}\\
		{\includegraphics[trim={0.5cm 0.0cm 0.5cm 1.0cm},clip ,
			width=0.99\columnwidth]{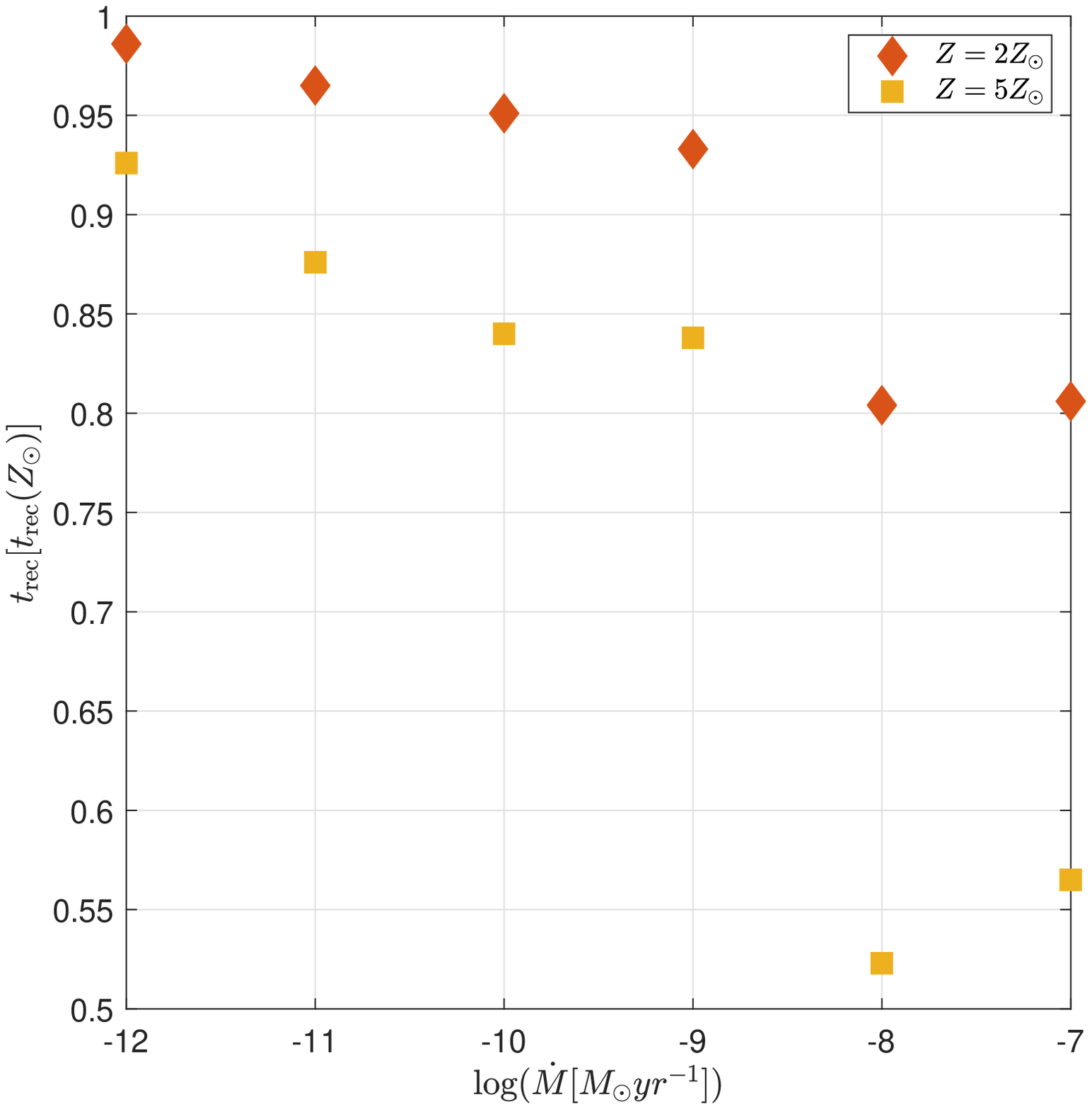}}
		\caption{Recurrence time vs. accretion rate for the three abundances, Top: Solar abundance; bottom: twice and five times the solar abundance.}\label{fig:trec}
	\end{center}
\end{figure}

We show the recurrence time ($t\rm_{rec}$) for the $Z=Z_\odot$ model in Figure \ref{fig:trec} (top) as a baseline for comparison with the enriched models, showing about one order of magnitude decrease in $t\rm_{rec}$ for one order of magnitude increase in $\dot{M}$. This is as expected and in agreement with theory and previous modeling \cite[]{Prikov1995,Yaron2005}. The bottom panel of Figure \ref{fig:trec} shows $t\rm_{rec}$ for the enriched models, in units of $t\rm_{rec}$ of the baseline model, showing a general decrease in $t\rm_{rec}$ with increase in $Z$. However, for high accretion rates ($\dot{M}\ge10^{-8}\dot{M} yr^{-1}$) the difference is more pronounced than for the lower $\dot{M}$s, and for the models with a higher enrichment ($Z=5Z_\odot$) the difference is more pronounced than for the less enriched models ($Z=2Z_\odot$). This means that for a system that accretes at a low rate, even an abundance of five times the solar abundance will reduce the recurrence time by only up to $\sim15\%$ whereas the same high abundance will reduce the recurrence period to about \textit{half} the time for high accretion rates, and by $\sim20\%$ for an abundance of $2Z_\odot$.
The reason that the accretion rate has such a stark influence on the extent of the affect of the enrichment, is due to the low accretion rates having another source of enrichment --- \textit{time}. A longer accretion phase allows more time for heavy elements to mix into the envelope from the underlying core, thus, externally adding heavy elements does not inflict as much of a relative enrichment as for systems with a shorter accretion phase, i.e., with a higher accretion rate. 

\begin{figure}
	\begin{center}
		{\includegraphics[trim={0.5cm 0.0cm 0.5cm 0.5cm},clip ,		width=0.99\columnwidth]{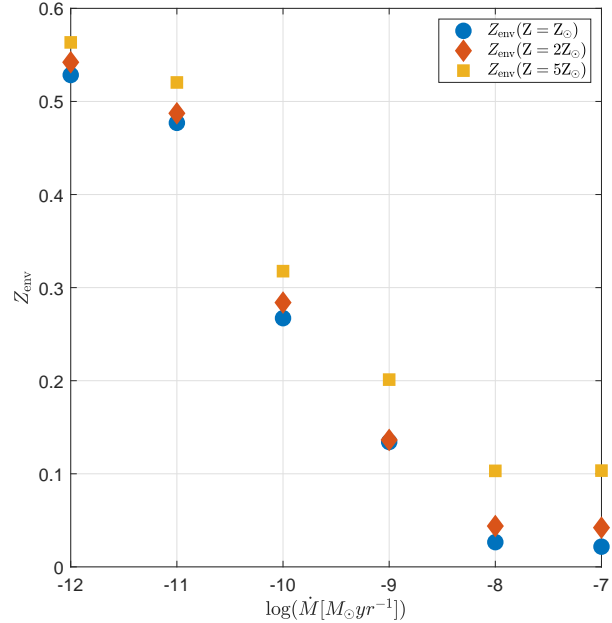}}
\caption{Envelope enrichment vs. accretion rate.}\label{fig:menv1}
	\end{center}
\end{figure}

\begin{figure*}
	\begin{center}
	{\includegraphics[trim={0.5cm 0.0cm 0.5cm 0.5cm},clip ,		width=0.99\columnwidth]{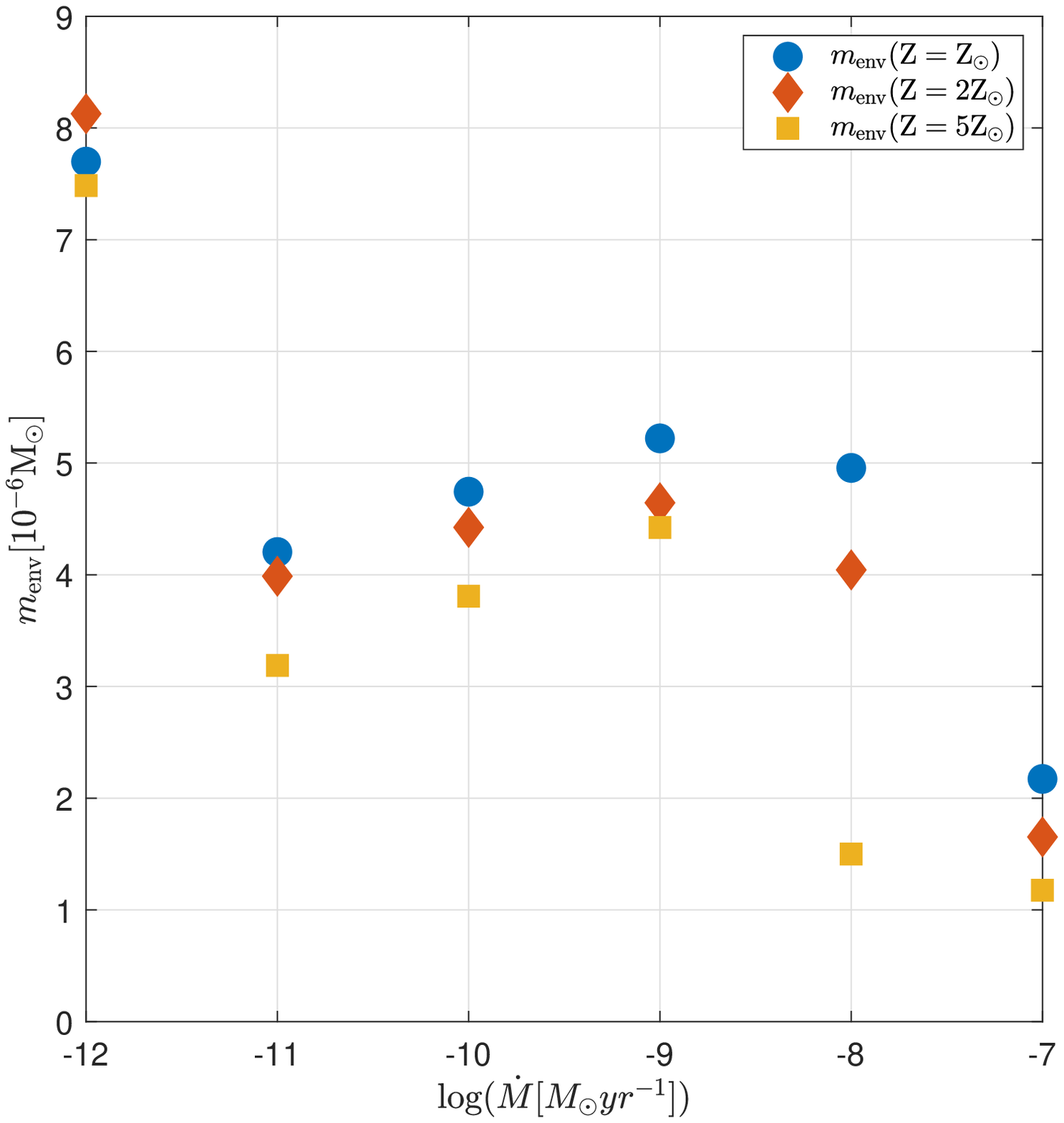}} {\includegraphics[trim={0.5cm 0.0cm 0.5cm 0.5cm},clip ,	width=0.99\columnwidth]{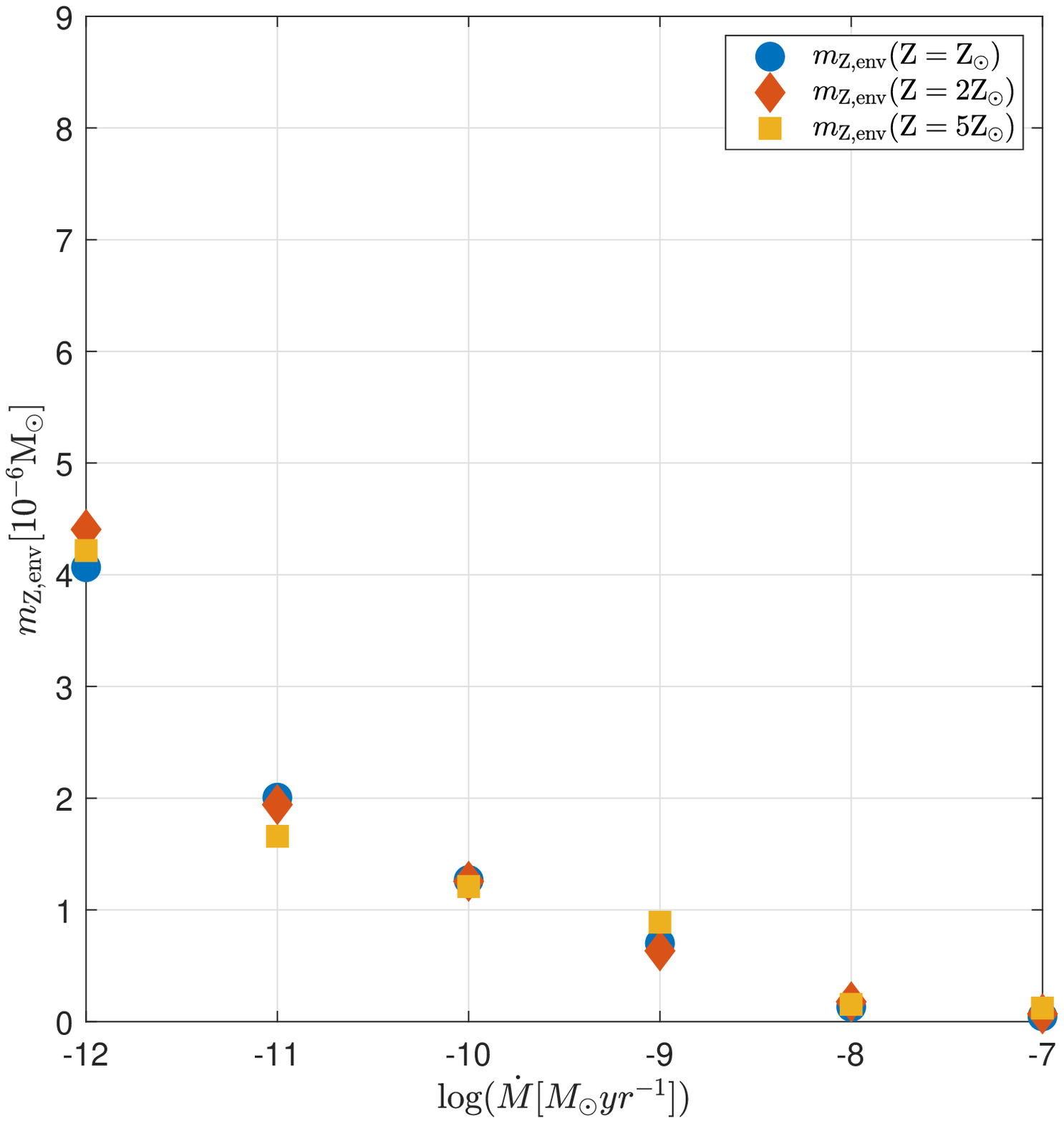}}
		\caption{Envelope total mass (left) and mass of heavy elements in envelope (right) vs. accretion rate.}\label{fig:menv2}
	\end{center}
\end{figure*}

\begin{figure*}
	\begin{center}
{\includegraphics[trim={0.5cm 0.0cm 0.5cm 0.5cm},clip ,	width=0.99\columnwidth]{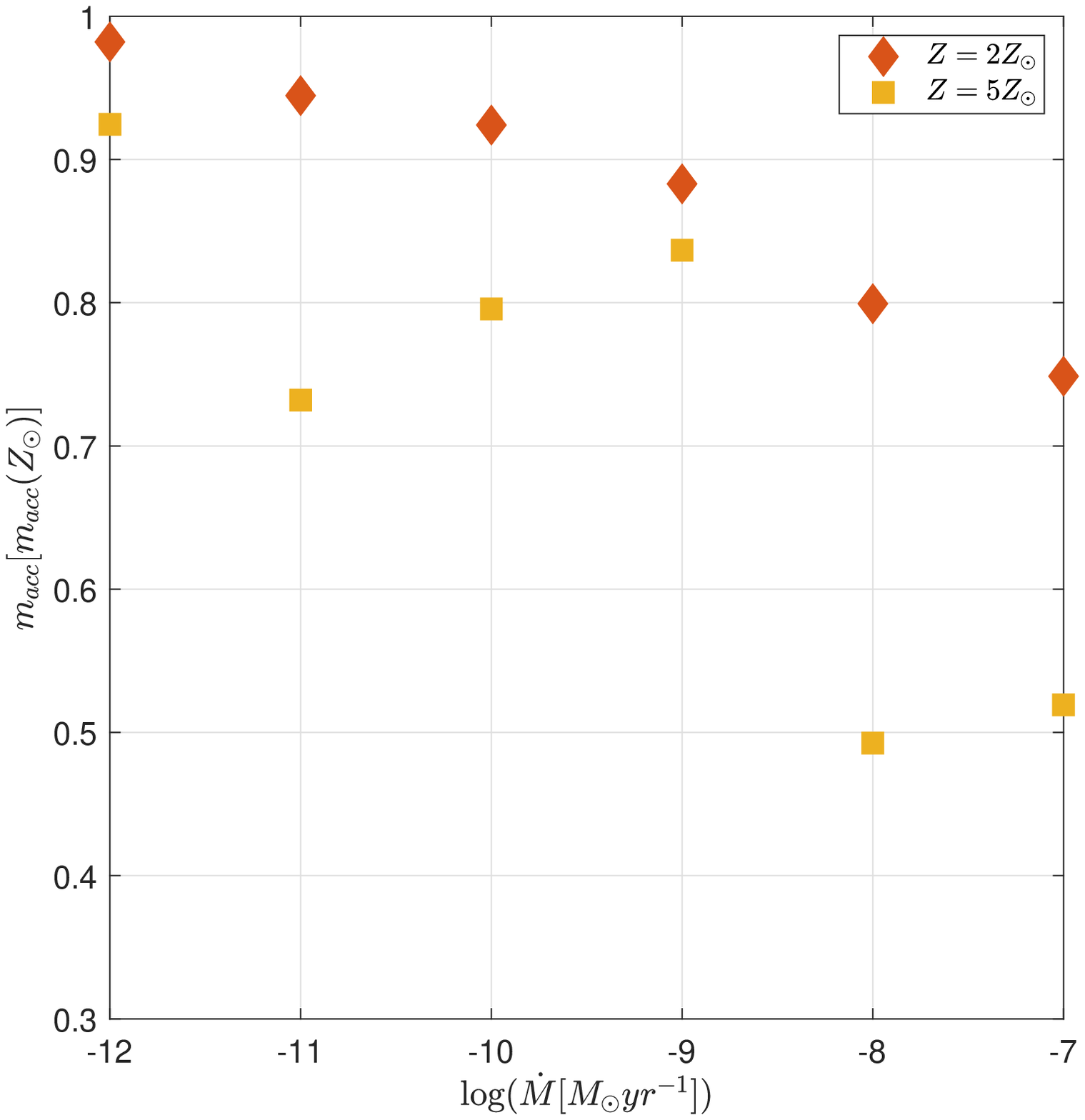}}
{\includegraphics[trim={0.5cm 0.0cm 0.5cm 0.5cm},clip ,	width=0.99\columnwidth]{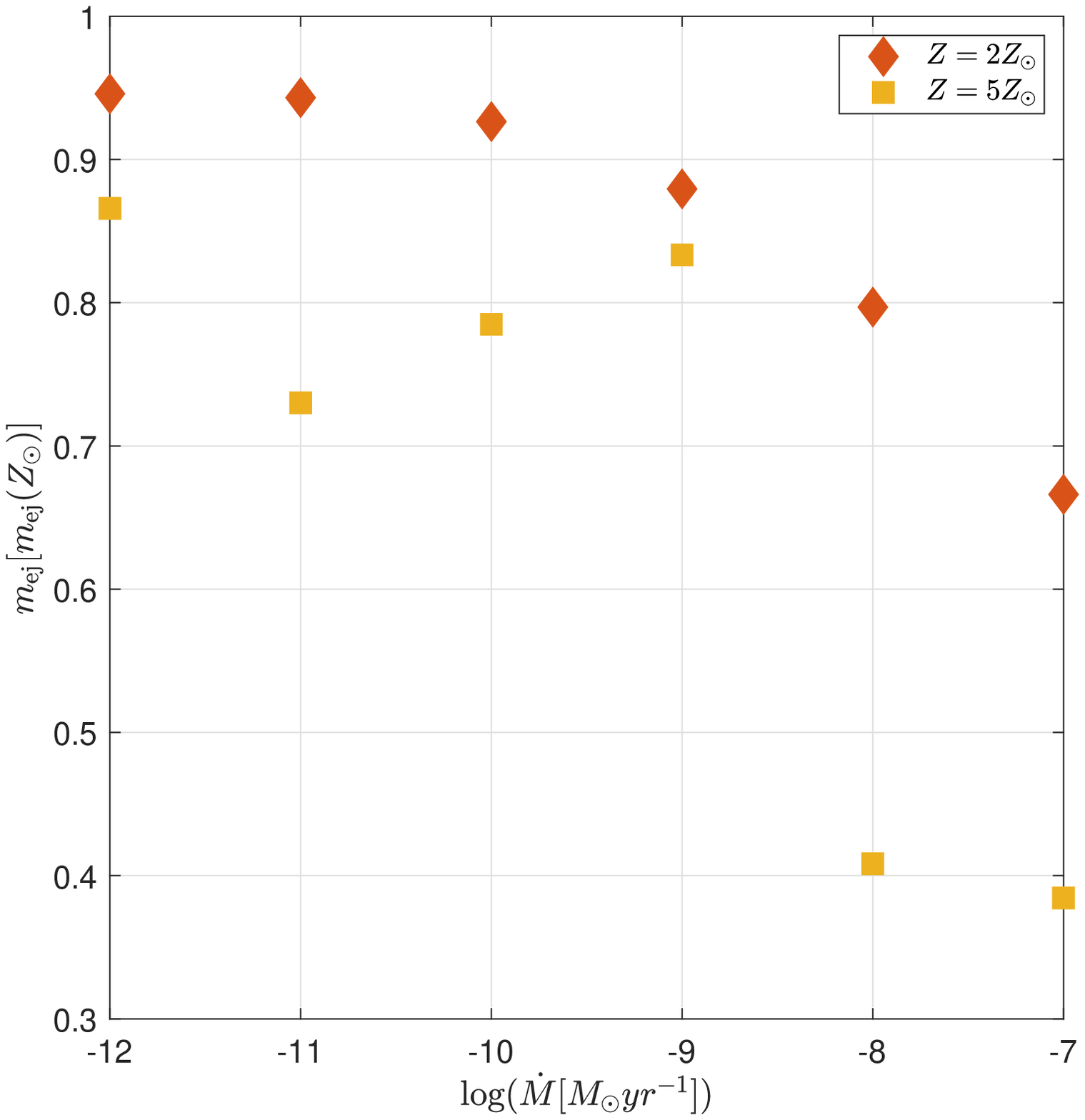}}
\caption{The accreted mass ($m\rm_{acc}$, left) and ejected mass ($m\rm_{ej}$, right) in units of $m\rm_{acc}$ and $m\rm_{ej}$ of the $Z=Z_\odot$ models vs. accretion rate.}\label{fig:macc_over_macc_1Z}
	\end{center}
\end{figure*}

This is further demonstrated via the envelope enrichment ($Z\rm_{env}$) at the end of each cycle, shown in Figure \ref{fig:menv1}. There is a distinct lower concentration of heavy elements in the WDs envelopes of models with higher accretion rates that is consistent for all three abundances (the two enriched models and the baseline solar abundance model). This plot additionally shows a consistently relative higher $Z\rm_{env}$ with higher enrichment of the accreted matter, and with the increase of the accretion rate. On the other hand, examining the left panel of Figure \ref{fig:menv2} reveals what seems like an opposite trend --- higher accretion rates generally have less massive envelopes for a higher $Z$ in the accreted matter.
This discrepancy is settled when considering, the actual \textit{amount} of heavy elements in the envelope, as shown in the right panel of Figure \ref{fig:menv2}. The difference between the curves of different enrichments almost vanishes. This means that the TNR occurs \textit{not for a critical amount of accreted mass, but for a  critical amount of accreted $Z$}. Of course, this means that for accreted matter with a higher enrichment, the \textit{total} accreted mass ($m\rm_{acc}$) will be smaller, as mentioned earlier. This is clear from Figure \ref{fig:macc_over_macc_1Z}  (left panel) which shows $m\rm_{acc}$ relative to the accreted mass of the baseline (solar abundance) models ($m{\rm_{acc}}(Z_\odot)$)  --- always below '1' and lower for $Z=5Z_\odot$ than for $Z=2Z_\odot$ --- and in agreement with the recurrence period (Figure \ref{fig:trec}) being shorter for the more enriched models. 

The ejected mass (Figure \ref{fig:macc_over_macc_1Z}, right panel) shows a similar behavior, for a similar reason --- less time to accrete means less time for mixing of matter from deeper layers into the accreted envelope, thus the ignition occurs at a shallower point, ending in less ejected mass. This further demonstrates that the amount of accreted Z --- rather than the total amount of accreted mass ---  is the key value that determines the development of a nova, for a given accretion rate.


\begin{figure*}
	\begin{center}
		{\includegraphics[trim={0.5cm 0.0cm 1.5cm 0.5cm},clip ,
			width=0.68\columnwidth]{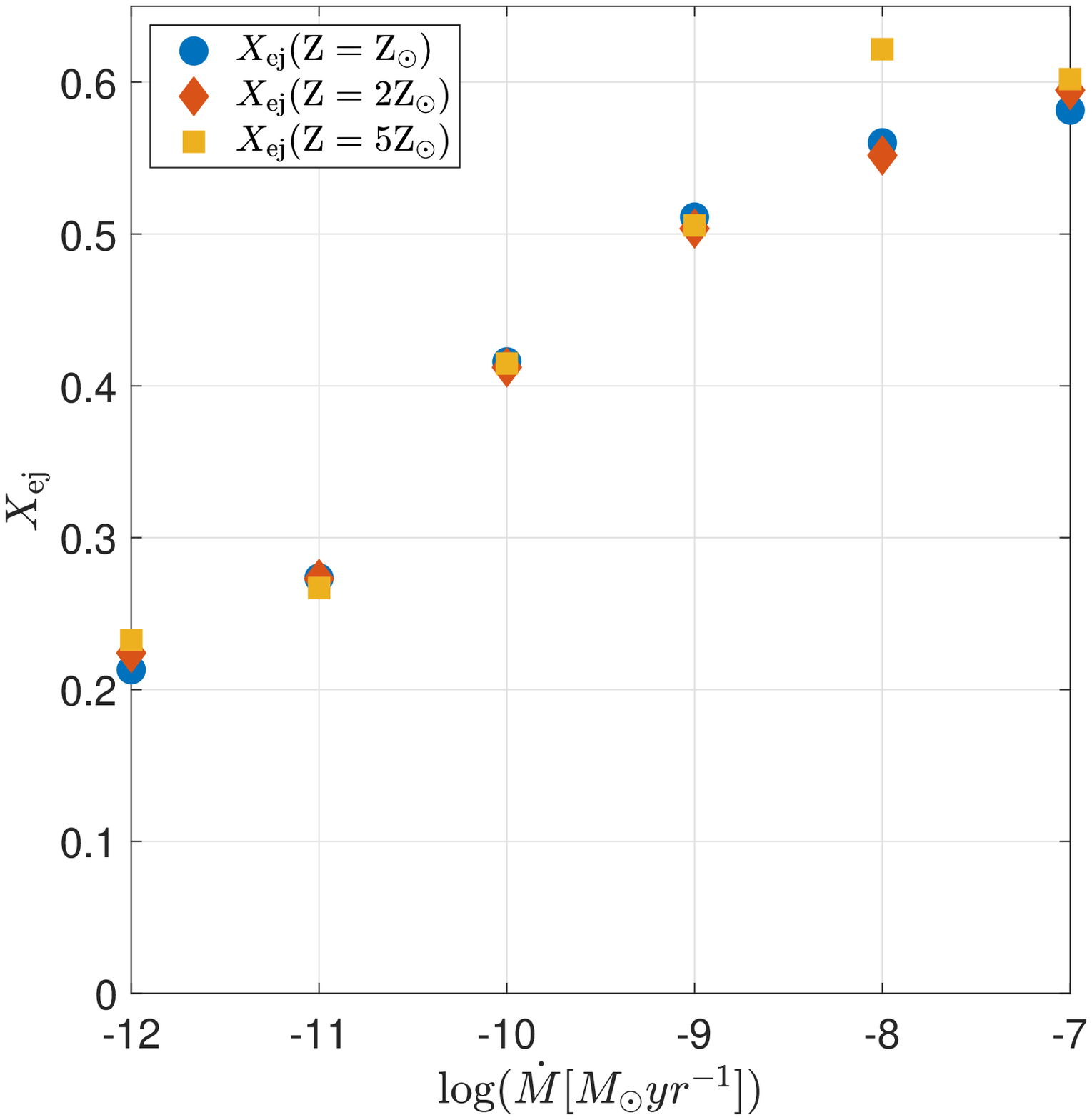}}
		{\includegraphics[trim={0.5cm 0.0cm 1.5cm 0.5cm},clip ,
			width=0.68\columnwidth]{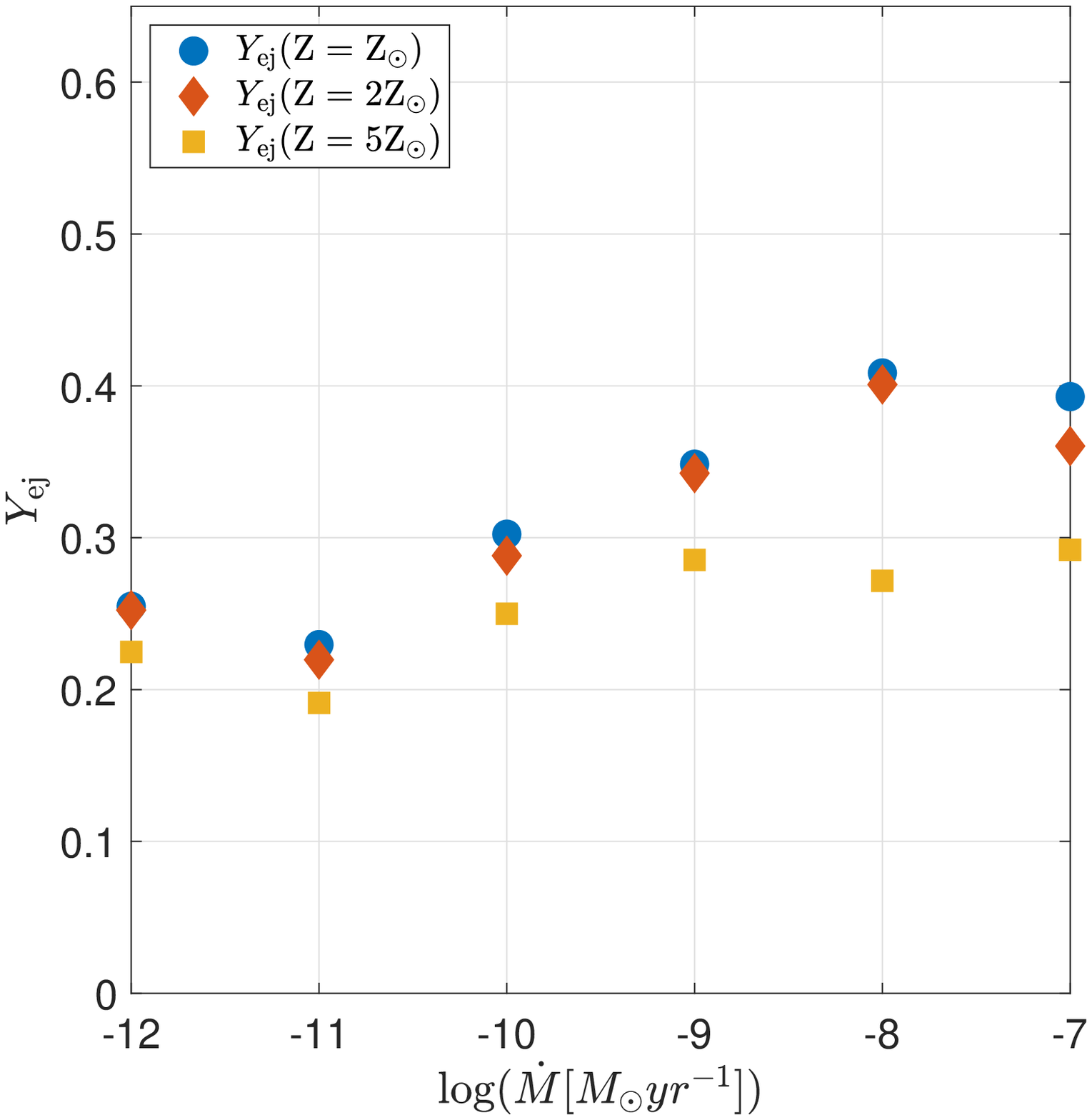}}
		{\includegraphics[trim={0.5cm 0.0cm 1.5cm 0.5cm},clip ,
			width=0.68\columnwidth]{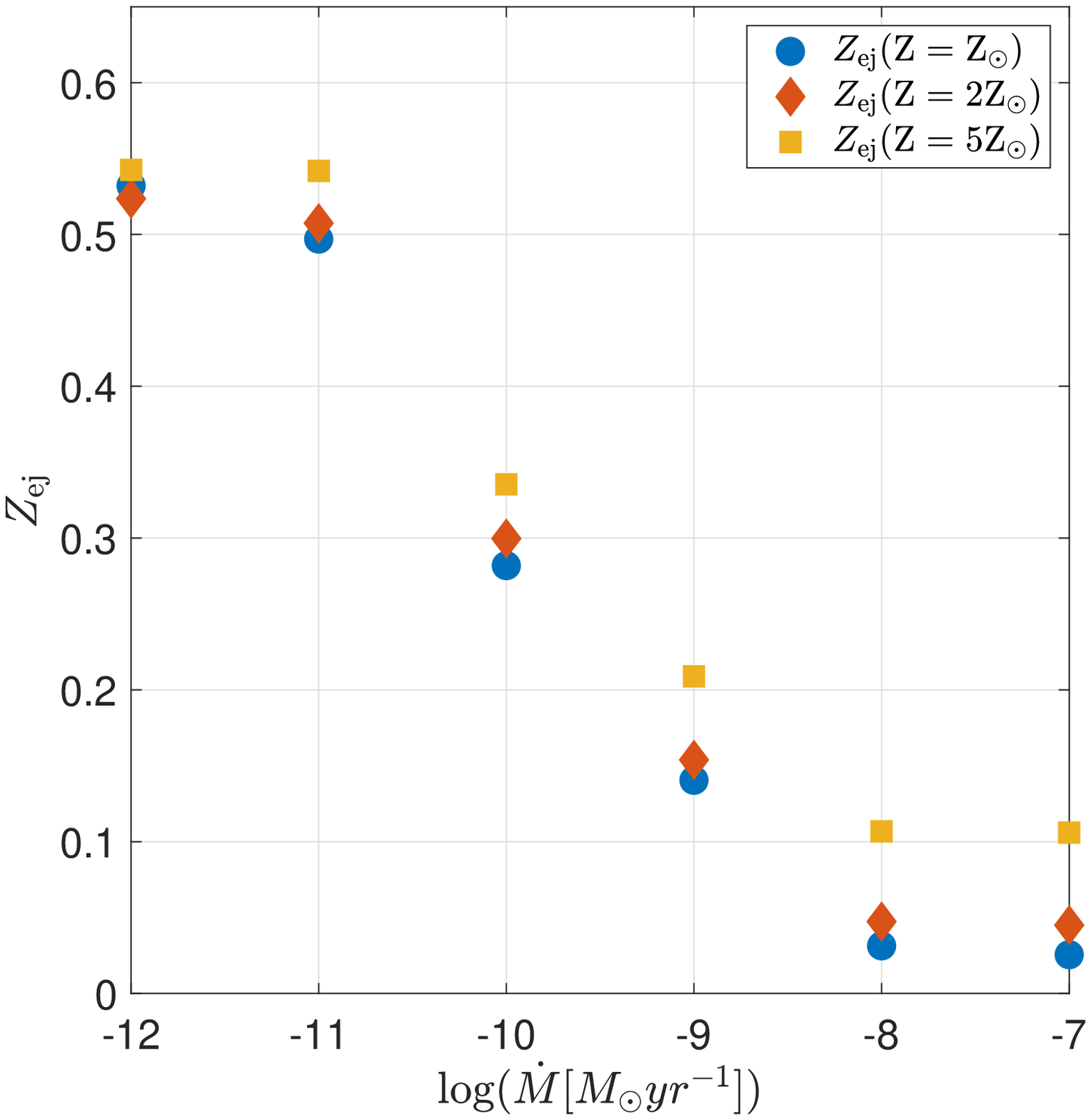}}
	\caption{Mass fractions of the ejected mass vs. accretion rate. Left: hydrogen ($X{\rm_{ej}}$); middle: helium ($Y{\rm_{ej}}$); and right: heavy elements ($Z{\rm_{ej}}$). where $X{\rm_{ej}}+Y{\rm_{ej}}+Z{\rm_{ej}}=1$.}\label{fig:XYZ}
	\end{center}
\end{figure*}

Examining the composition of the ejected mass (Figure \ref{fig:XYZ}) shows more hydrogen (left panel) for higher accretion rates; a smaller slope, yet still generally more helium (middle panel) for higher accretion rates as well; and a strong opposite trend for the heavy element content (right panel). This is all in excellent agreement with previous solar abundant accreted matter models \cite[]{Prikov1995, Yaron2005}. It is also clear from Figure \ref{fig:XYZ} that the hydrogen content is hardly affected by changing the heavy element enrichment of the donor, while the helium content is reduced with enrichment, and the heavy element content is evidently increased. Of course we must consider that this is a direct result of the accreted matter being enriched in that exact such manner, however the relation is not uniform for all the models, but rather we find (1) a dependence on the accretion rate --- the difference in ejecta abundance between the different initial enrichments is more substantial for higher accretion rates --- as found in other parameters discussed earlier, and (2) for the high accretion rates, the relation between the ejecta abundances of the enriched models and the base models \textit{is} close to the initial enrichment relation of the accreted mass --- roughly twice and five times the amount of $Z$ in the corresponding models. However, as the accretion rate decreases, the differences are reduced, and for our lowest accretion rate the differences are negligible, demonstrating once again, that enriching the accreted material has a non-negligible affect only on high accretion rates.

\begin{figure}
	\begin{center}
		{\includegraphics[trim={0.5cm 0.0cm 0.5cm 0.5cm},clip ,			width=0.99\columnwidth]{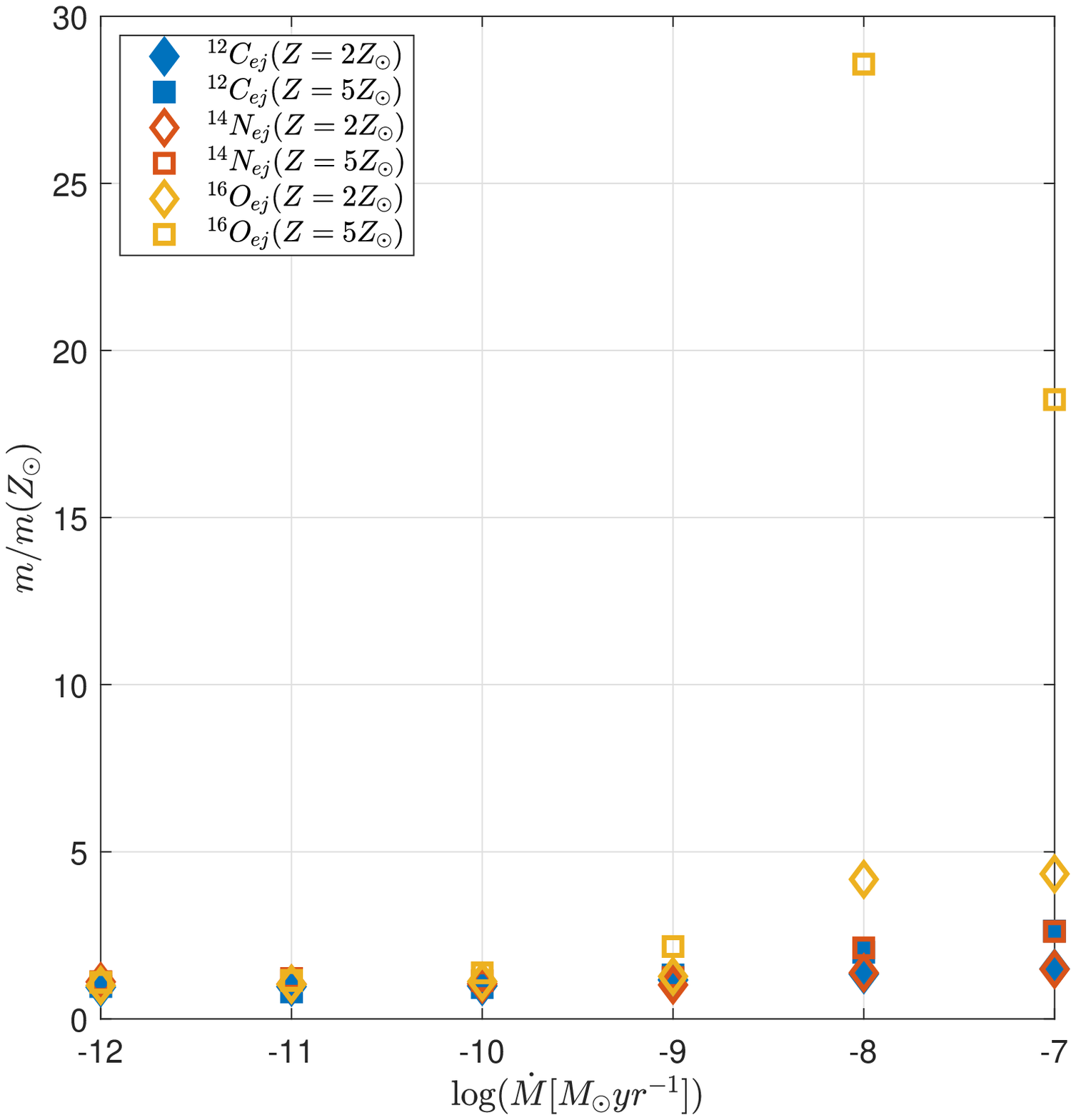}}		
		\caption{CNO mass fractions of the total $Z\rm_{ej}$ in units of CNO mass fractions of the total $Z\rm_{ej}$ of the base models ($Z=Z_\odot$) vs. accretion rate.}\label{fig:CNO}
	\end{center}
\end{figure}

\begin{figure}
	\begin{center}
		{\includegraphics[trim={0.5cm 0.0cm 0.5cm 0.5cm},clip ,
			width=0.99\columnwidth]{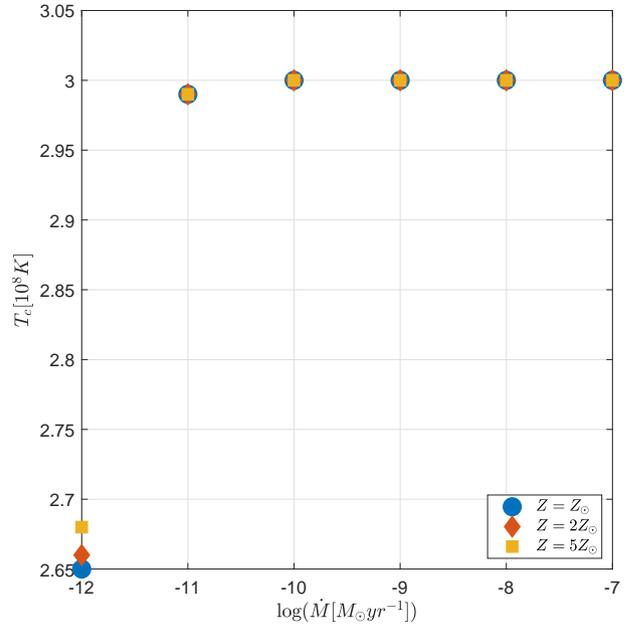}}
		\caption{The core temperature ($T\rm_{c}$) of the WD vs. accretion rate.}\label{fig:Tc}
	\end{center}
\end{figure}

\begin{figure}
	\begin{center}
		{\includegraphics[trim={0.5cm 0.0cm 0.5cm 0.5cm},clip ,
			width=0.99\columnwidth]{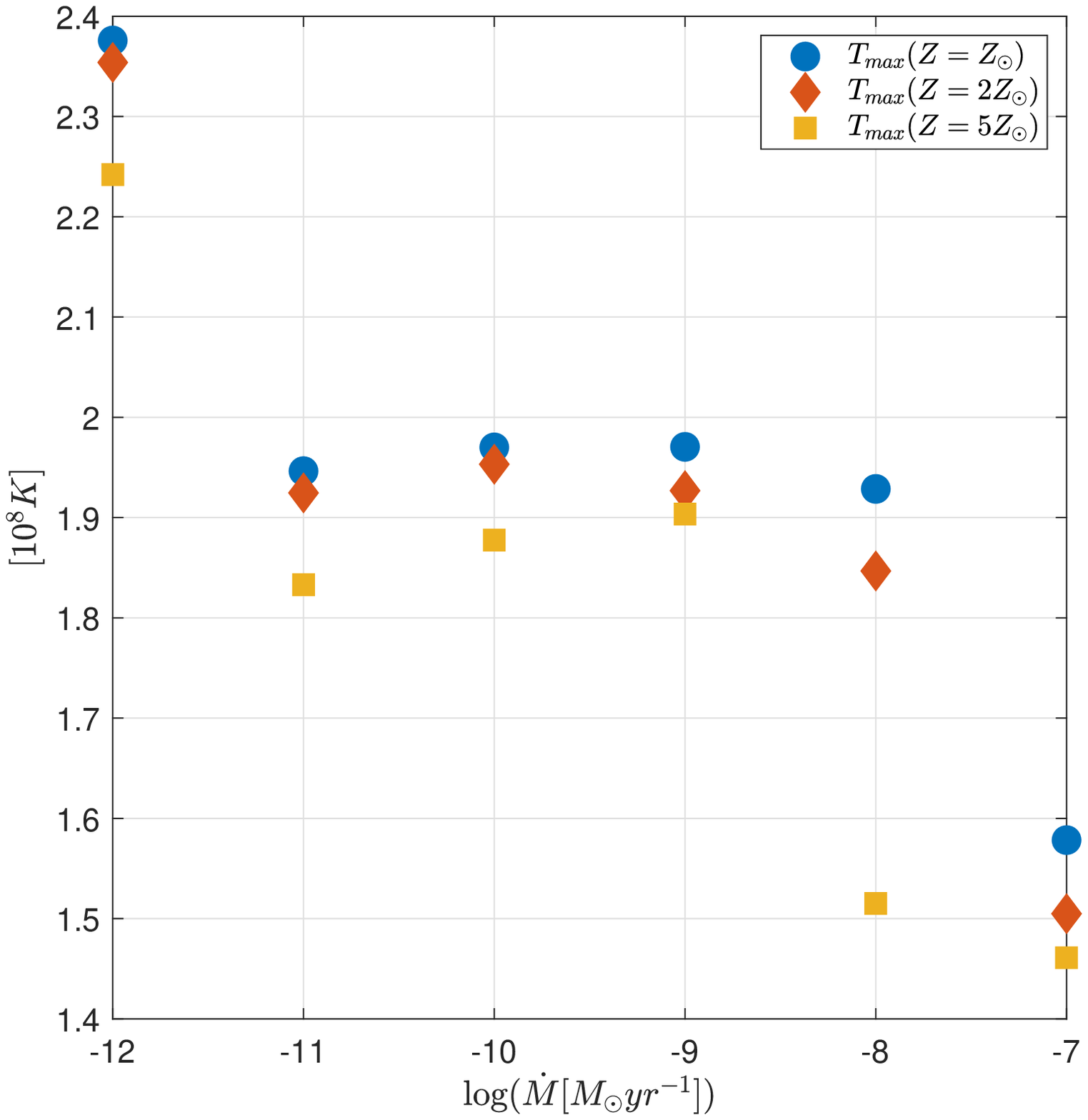}}
	\caption{The maximum temperature ($T\rm_{max}$) reached during a nova eruption vs. accretion rate.}\label{fig:Tmax}
	\end{center}
\end{figure}

Looking deeper into the composition of $Z\rm_{ej}$ (Figure \ref{fig:CNO}) reveals that the enrichment of the accreted matter has a minuscule affect on the carbon and nitrogen abundances in the ejecta, reaching at the most two to three times the abundance of the baseline ($Z=Z_\odot$) models for the $5Z_\odot$, highest accretion rate models. The oxygen abundance for low accretion rates is un-affected in the same manner, regardless of the enrichment, however, in stark contrast, the oxygen abundance in the ejecta of the $2Z_\odot$ models accreting at high rates is nearly four times the abundance of the corresponding baseline models, and an order of magnitude higher for the $5Z_\odot$ models. This is an important result, since it may possibly be confronted with observations.  

We additionally examined the temperature of the WD in three places --- the core ($T_c$), the surface (effective temperature, $T\rm_{eff}$), and the maximum temperature ($T\rm_{max}$) in the envelope. The core temperature seems to show no difference at all between the enrichments, except for our lowest accretion rate. This may be understood by recalling the time scales of the recurrence times (Fig. \ref{fig:trec}). Our lowest accretion rate, for solar abundance, accretes quiescently for a few million years before erupting, allowing enough time for the core to cool between eruptions, reaching the beginning of an eruption at a lower temperature than the beginning of the previous cycle. For shorter accretion phases (i.e., higher accretion rates) after an eruption is quenched, as the core begins to cool, an eruption heats the WD from the outside, and the heat slowly diffuses to the deeper layers. For very high accretion rates, over hundreds or thousands of cycles (or more) the heat from the eruptions will eventually diffuse all the way to the center of the WD, heating it as well \cite[]{Epelstain2007,Hillman2016,Hillman2020a}, however the study here included tens of cycles so we do not observe this long term heating. The difference we find in the $T\rm_c$ of the different enrichments for our lowest accretion rate ($10^{-12}M_\odot yr^{-1}$) is a direct result of the shorter time needed to achieve the critical heavy element mass --- the more enriched models required less time, thus had less time to cool. Noting that Figure \ref{fig:trec} shows very minor differences in $t\rm_{rec}$ for the different enrichments, at this accretion rate, a few percent of the accretion time is sufficient to have an effect on the cooling.

\begin{figure}
	\begin{center}
{\includegraphics[trim={0.5cm 0.0cm 0.5cm -0.5cm},clip ,			width=0.99\columnwidth]{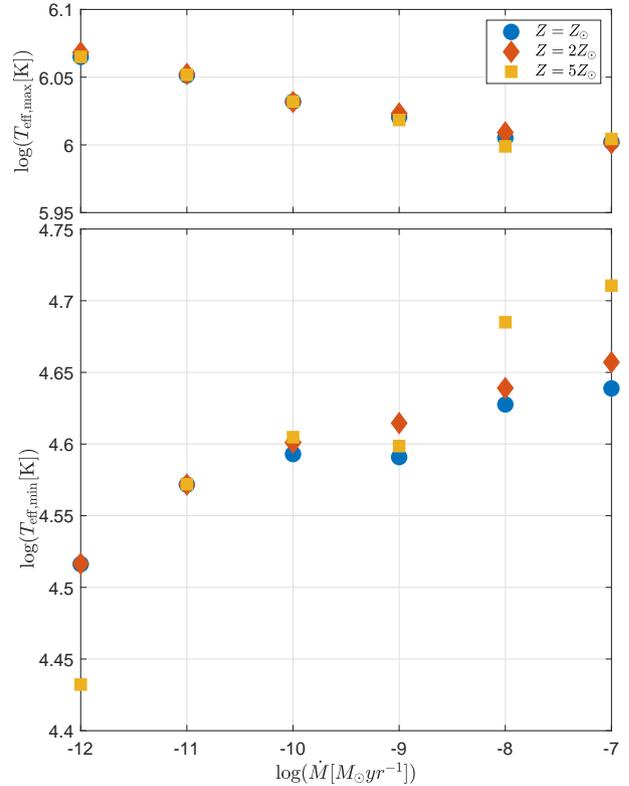}}
		\caption{The effective temperature ($T\rm_{eff}$) vs. accretion rate. Top: maximum; bottom: minimum.}\label{fig:Teff}
	\end{center}
\end{figure}

In contrast with the nearly indistinguishable differences in $T_c$, the maximum temperature ($T\rm_{max}$) attained during an eruption (Figure \ref{fig:Tmax}) varies from as high as nearly $2.4\times10^8K$ for $\dot{M}=10^{-12}M_\odot yr^{-1}$ to lower than $1.5\times10^8K$ for $\dot{M}=10^{-7}M_\odot yr^{-1}$. As we found for many of the other features, the variations due to change in accretion rate are more substantial than the differences due to enrichment, and the results of our solar abundant models are in excellent agreement with previous modeling \cite[]{Prikov1995, Yaron2005}. Examining the different enrichments, we see a consistently lower $T\rm_{max}$ for models with more enriched accreted matter. The reason for this is because a higher enrichment allows the nuclear reactions in the CNO cycle to occur at a faster rate, meaning that the temperature does not need to rise as much in order to initiate a TNR, resulting in a maximum temperature that is slightly lower for the $Z=2Z_\odot$ models than for their corresponding $Z=Z_\odot$ baseline models and even lower for the $Z=5Z_\odot$ models, consistent for the entire range of accretion rates. 

in Figure \ref{fig:Teff} we show the minimum and maximum effective temperatures ($T\rm_{eff}$) attained during a nova, starting from the TNR and until the accretion begins again. For the maximal $T\rm_{eff}$ ($T\rm_{eff,max}$) we find that all six models of each $\dot{M}$ exhibit virtually identical values, regardless of the enrichment. This is because the maximum effective temperature is attained right before and right after the WD expels mass, i.e., during the nova, when the radius it at its smallest. Since the luminosity (which is limited to the Eddington luminosity) and the minimum radius are both determined by the WD mass, $T\rm_{eff,max}$ should be constant regardless of the amount of ejected mass (or heavy elements) and therefore of the accretion rate. The slight difference between rates is a result of the accretion luminosity being proportional to the accretion rate, thus for higher rates the accreted envelope is forced to expand somewhat more than than for lower rates, resulting in a somewhat larger radius and thus a slightly lower $T\rm_{eff,max}$. 

The minimum effective temperature ($T\rm_{eff,min}$) increases with increasing $\dot{M}$ more substantially than the decrease in $T\rm_{eff,max}$. This is because the minimum effective temperature is obtained when the WD is losing mass and has expanded to its largest radius. Since less accreted mass results in a weaker eruption the expansion is less pronounced \cite[]{Yaron2005,Hillman2019} resulting in a higher $T\rm_{eff,min}$ for a higher $\dot{M}$. Additionally, and more importantly for the focus of this paper, the $T\rm_{eff,min}$ shows a \textit{clear} difference between the enrichments --- more enriched accreted matter results in a steeper slope, while for our lowest $\dot{M}$ a higher enrichment yields a lower $T\rm_{eff,min}$, gradually changing until for our highest $\dot{M}$ the $T\rm_{eff,min}$ is the highest for the most enriched accreted matter. This means that the connection between accretion rate and effective temperature during mass loss (i.e., radial expansion) is stronger when the accreted matter is more enriched. The presence of more heavy elements cause the low rates, for which expansion is maximum, to expand even more, and for high rates, for which the expansion was minimal, to expand less. The enrichment accentuates this behavior.

\begin{figure}
	\begin{center}
	{\includegraphics[trim={0.5cm 0.0cm 0.2cm 0.1cm},clip ,width=0.99\columnwidth]{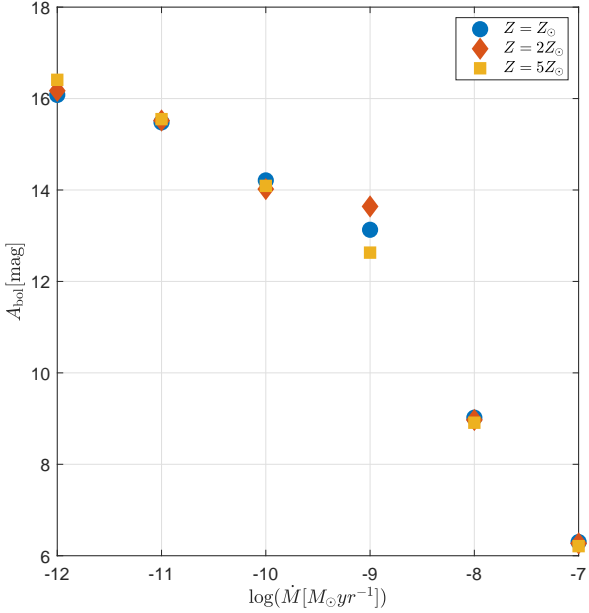}}
		\caption{Bolometric eruption amplitudes vs. accretion rate.}\label{fig:Amplitudes}
	\end{center}
\end{figure}

The bolometric amplitude of eruption ($A\rm_{bol}$), as may be seen in Figure \ref{fig:Amplitudes}, shows that the enrichment of the accreted matter has a negligible affect on it, while it decreases drastically with increasing $\dot{M}$. This is as a result of the same mechanism that determined the minimum effective temperature --- less time to cool means a higher minimum magnitude before eruption. This, together with the maximum magnitude being limited by $L\rm_{edd}$, yields a smaller eruption amplitude for higher accretion rates.\footnote{The differences in $T\rm_{eff,min}$ for the different enrichments do not show up in the amplitude because on a logarithmic scale they are very small.}.
 
The amplitude is one of the few observable features of a nova and therefore models such as these \cite[and others, e.g.,][]{Yaron2005,Shara2018} can be used to identify the average mass accretion rate of a system prior to an observed eruption. Unfortunately, the results here clearly show that the amplitude cannot assist in determining the composition of the accreted matter. 

\begin{figure}
\begin{center}
{\includegraphics[trim={0.55cm 1.6cm 0.15cm 0.1cm},clip ,width=0.99\columnwidth]{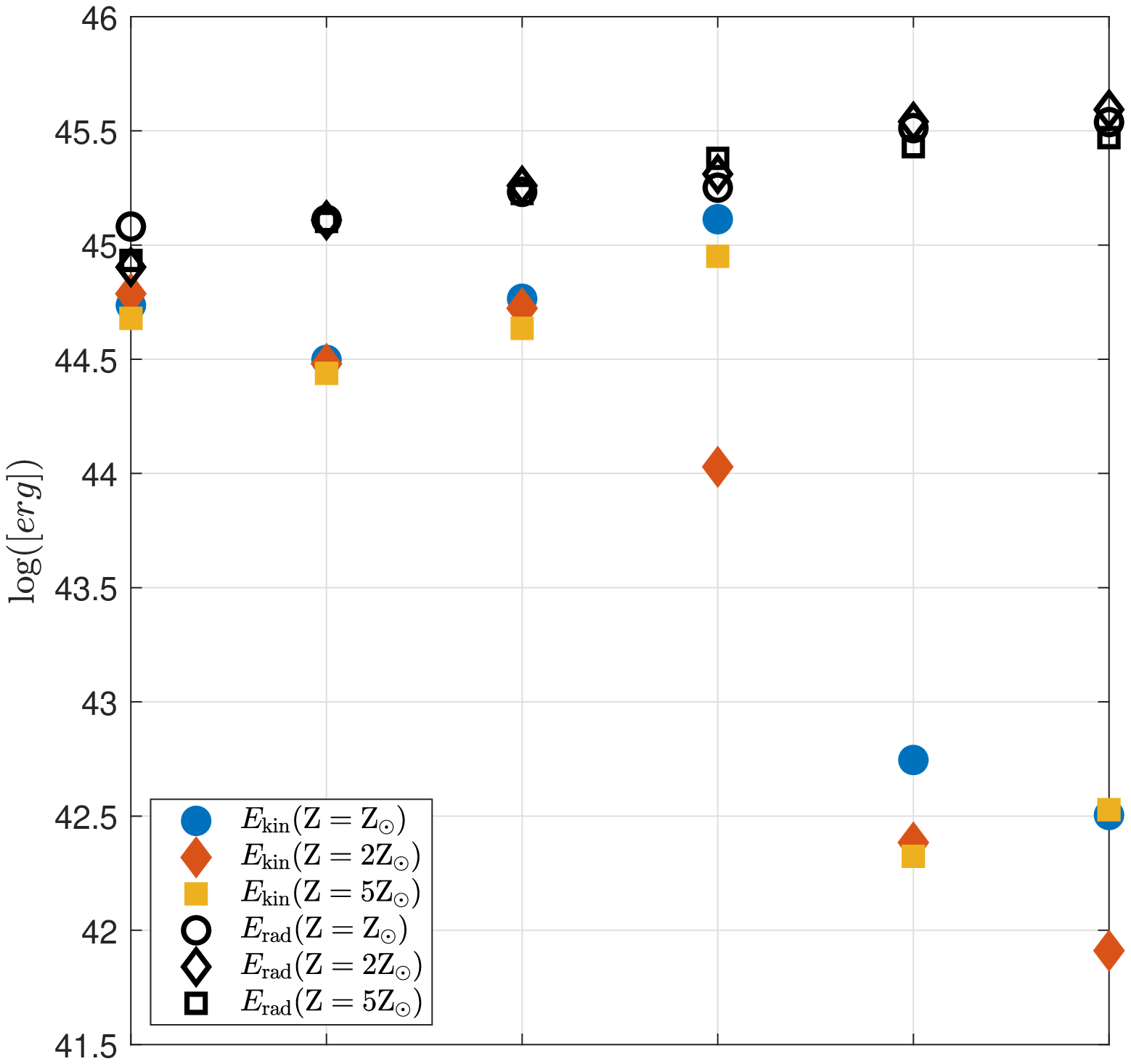}}
{\includegraphics[trim={0.5cm 0.0cm 0.2cm 0.8cm},clip ,width=0.99\columnwidth]{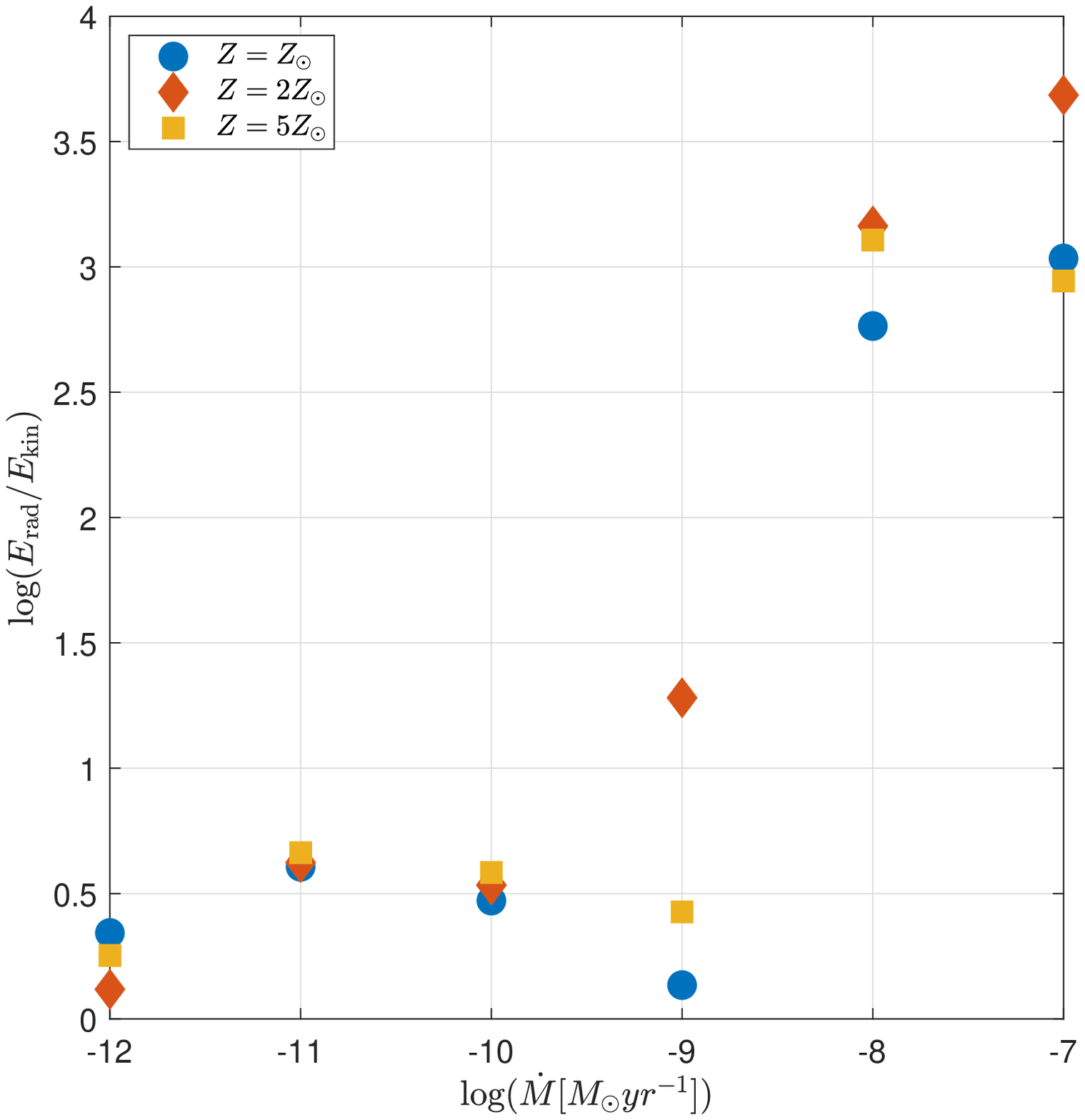}}
\caption{Radiative and kinetic energies (top) and the ratio of these energies (bottom) vs. accretion rate.}\label{fig:Energy}
\end{center}
\end{figure}

The nuclear energy produced in a single nova eruption is the product of the energy released by the fusion of four protons into helium via the CNO cycle and the amount of hydrogen in mass that was burnt. This yields a few times $10^{45}$erg for the models used here, which is in agreement with many estimates, e.g., \cite{Pay1957,Bath1978} and \cite{Mukai2019}. This energy is the source of the gravitational lift, the radiated energy ($E\rm_{rad}$) and the kinetic energy ($E\rm_{kin}$). The ratio $E\rm_{rad}/E\rm_{kin}$ has been examined and discussed in \cite{Shara2010a} and found to vary immensely with $\dot{M}$. $E\rm_{rad}$ is calculated by integrating the bolometric luminosity over the eruption time, which, during eruption, is roughly the Eddington luminosity, that is, a few times $10^4L_\odot$ and proportional to the WD mass, making $E\rm_{rad}$ roughly the same for all the models\footnote{The Eddington luminosity is also inversely proportional to the opacity which varies during a cycle and between cycles, which is what causes the variations between the models (Figure \ref{fig:Energy})}. On the other hand, the velocity varies greatly with $\dot{M}$ and is correlated with $m\rm_{ej}$ \cite[]{Yaron2005}. This is because the models that accrete at a higher rate, eject less mass, thus, they need less energy to lift $m\rm_{ej}$ gravitationally, resulting in a slower velocity. $E\rm_{rad}$ and $E\rm_{kin}$ are shown in Figure \ref{fig:Energy}, as well as the ratio $E\rm_{rad}/E\rm_{kin}$ varying 3-4 orders of magnitude --- higher for higher accretion rates --- which is in agreement with \cite{Shara2010a}. The ratio also shows a variation of about one order of magnitude between different enrichments for high $\dot{M}$s, although since multiple parameters are involved in producing these energy values (velocity, ejected mass, amount of CNO, bolometric luminosity, eruption time) the resulting energy ratio differences between enrichments \textit{seems} sporadic but is simply the result of all these parameters affecting the energy budget in different directions.

\section{Discussion and conclusions}\label{sec:conclusions}
We have carried out a series of nova simulations with different accretion rates and enrichments in order to assess the influence of the accreted matter (i.e., the donor envelope) being enriched with heavy elements, on the resulting nova eruptions. The goal of this study ultimately being to determine if enriching the accreted material would show, in any of the resulting parameters, a difference substantial enough to differentiate observationally. 

We found a significant difference in the time between eruptions for high accretion rates, i.e., for models in the regime of recurrent novae ($\sim10^{-8}-10^{-7}M_\odot yr^{-1}$).  To demonstrate the impact of this for practical purposes, we examine as an example, the wind model in \cite{Hillman2021} during its epoch of an average $\dot{M}$ of $\sim6\times10^{-8}M_\odot yr^{-1}$ for which the recurrence time is $\sim25$ years. If the accreted matter were to have twice the amount of heavy elements relative to solar composition, the recurrence period would be $\sim80\%$ of what they obtained, i.e., $\sim20$ years rather than $\sim25$. This is an important result, although if it were an actual observed RN with an actual $t\rm_{rec}\approx20$ years, it \textit{could} be attributed to a WD mass of 1.25$M_\odot$ that accretes matter enriched to twice the solar composition, but due to the nature of the dependence of $t\rm_{rec}$ on the WD mass \cite[see eq. 5 and 6 in][]{Hillman2016} it \textit{could also} be attributed to a \textit{slightly} more massive WD accreting solar composition matter (not more than a few percent more massive, even if we were to consider the five fold enrichment). 
This means that the time between eruptions, although substantially different for some of the models, cannot be used alone as an observational tool to determine the composition of the donor envelope. Another example may be the estimated modeled accretion rate for the classical nova V5589 Sgr in \cite{Eyres2017}, which is based on solar composition accreted material. They state that since the donor in the system is a sub-giant, the accreted material may be somewhat enriched, and thus the estimated accretion rate may be somewhat different. Here we see that the difference would be small for two reasons (a) it is in the low $\dot{M}$ regime and (b) the enrichment of a sub-giant's envelope is lower than the levels examined here, thus their estimated accretion rate would not be altered from the results here, also meaning that the outcome of a nova with a sub-giant donor would be the same as for a MS donor, regardless of the accretion rate. In fact, we note that exploring the high enrichment of five time the solar abundance is an academic case and to our best knowledge even evolved AGB stars do not become so enriched.

Most of the other parameters, namely, the different temperatures, eruption amplitude and energy balance as well as the envelope, the accreted and ejected masses, and their heavy element content,  showed minor variations between enrichments, except for one feature --- the oxygen in the ejecta for high $\dot{M}s$ --- which showed a substantial increase with enrichment, not only for the $Z=5Z_\odot$ models, but also for the $Z=2Z_\odot$ models. Its presence in the ejected material shows a strong dependence --- much larger than its relevant enrichment --- on the level of enrichment for the RN regime. This means that observations of an unusually large amount of $^{16}O$ in the ejecta (expanding shell) would imply that the accreted matter (i.e., the donor envelope) was more enriched than solar abundance, possibly pointing to (or supporting) the donor being a highly evolved star --- an AGB. However, in such systems, the WD is often embedded in (and/or contaminated by emission from) the envelope of the giant. 

The CNO ratio of symbiotic systems show a wide spread of values \cite[see table 1 in][]{Nussbaumer1988}, so extracting CNO ratios of the ejected material alone from the nova eruption is complex for these systems, making this high mass fraction of oxygen challenging to confront at this point with observations of systems hosting an evolved, giant donor. Moreover, given that the assumption that the wind of a giant is enriched to twice the solar abundance is a high estimate, we conclude that enriched accreted material probably will not have a \textit{detectable} affect on the observed composition of an expanding shell. 

Nonetheless, the understanding that for RN systems with a donor envelope of highly enriched matter, the ejected material \textit{will be} enriched in oxygen at a substantially higher level than its initial enrichment in the accreted material, stands as one of the main results of this study, the key result being the realization that it is not a critical amount of mass, but a critical amount of heavy elements that will trigger a TNR on the surface of a WD.

\section{Data availability}
The data underlying this article will be shared on reasonable request to the corresponding author.

\bibliographystyle{aasjournal}

\end{document}